\documentclass[10pt, conference, compsocconf]{IEEEtran}

\ifCLASSINFOpdf
   \usepackage[pdftex]{graphicx}
   \graphicspath{{./figures/}}
\else
   \usepackage[dvips]{graphicx}
   \graphicspath{{./figures/}}
   \DeclareGraphicsExtensions{.eps}
\fi

\usepackage{comment}
\usepackage{amsmath}
\usepackage{multirow}
\usepackage{soul}
\usepackage{color}
\usepackage{color, colortbl}
\usepackage{subfig}
\usepackage{bm}
\definecolor{Gray}{gray}{0.9}
\definecolor{red}{rgb}{0.0, 0.0, 0.0}
\definecolor{r2}{rgb}{0.0, 0.0, 0.0}

\usepackage{url}

\begin{document}
%
\title{
ECG Heartbeat Classification: A Deep Transferable Representation
}
%
%
%

\author{
    \IEEEauthorblockN{Mohammad Kachuee, Shayan Fazeli, 
    Majid Sarrafzadeh}
    \IEEEauthorblockA{University of California, Los Angeles (UCLA)\\
    Los Angeles, USA}
}

\maketitle

\begin{abstract}


Electrocardiogram (ECG) can be reliably used as a measure to monitor the functionality of the cardiovascular system. Recently, there has been a great attention towards accurate categorization of heartbeats. While there are many commonalities between different ECG conditions, the focus of most studies has been classifying a set of conditions on a dataset annotated for that task rather than learning and employing a transferable knowledge between different tasks. In this paper, we propose a method based on deep convolutional neural networks for the classification of heartbeats which is able to accurately classify five different arrhythmias in accordance with the AAMI EC57 standard. Furthermore, we suggest a method for transferring the knowledge acquired on this task to the myocardial infarction (MI) classification task. We evaluated the proposed method on PhysionNet's MIT-BIH and PTB Diagnostics datasets. According to the results, the suggested method is able to make predictions with the average accuracies of \bm{$93.4\%$} and \bm{$95.9\%$} on arrhythmia classification and MI classification, respectively.
\end{abstract}


%
\IEEEpeerreviewmaketitle

\section{Introduction}
\label{sec:Introduction}
%
ECG is widely used by cardiologists and medical practitioners for monitoring the cardiac health. The main problem with manual analysis of ECG signals, similar to many other time-series data, lies in difficulty of detecting and categorizing different waveforms and morphologies in the signal. For a human, this task is both extensively time-consuming and prone to errors. Note that the proper diagnosis of cardiovascular diseases is of paramount importance since these are the cause of death for about one-third of all deaths around the globe \cite{who2018}.
For instance, millions of people experience irregular heartbeats which can be lethal in some cases. Therefore, accurate and low-cost diagnosis of arrhythmic heartbeats is highly desirable \cite{society2018}.

To address the problems raised with the manual analysis of ECG signals, many studies in the literature explored using machine learning techniques to accurately detect the anomalies in the signal \cite{esmaili2017nonlinear,dastjerdi2017non}. Most of these approaches involve a preprocessing phase for preparing the signal (e.g., passing it through band-pass filters, etc). Afterwards, the handcrafted features which are mostly statistical summarizations of signal windows are extracted from these signals and used in further analysis for the final classification task. As for the inference engine, conventional machine learning approaches for ECG analysis include Support Vector Machines, multi-layer perceptrons, decision trees, etc.
\cite{inan2006robust,sayadi2010robust,kachuee2017cuffless}

These handcrafted features provide us with an acceptable representation of the signal, based on recent machine learning studies, automated feature extraction and representation methods are proven to be more scalable and are capable of making more accurate predictions. An end-to-end deep learning framework allows the machine to learn the features that are best suited to the specific task that it is dedicated to carry out \cite{Acharya2017,Kiranyaz2016,Jin2017}. This approach provides us with a more accurate representation of ECG signal using which the machine can compete with a human cardiologist in analyzing the signal \cite{rajpurkar2017cardiologist}. Deep learning approaches, however, contain a tremendously large amount of variables which require massive amounts of data to be trained. 

One way of dealing with the need to a massive amount of data is the concept of knowledge transfer between different tasks. In computer vision, as an example, ImageNet dataset along with the state of the art deep learning models have been used to transfer knowledge between different image understanding tasks \cite{Oquab2014LearningAT}. As another example, it has been shown that different sentence categorization tasks can share a considerable amount of sentence understanding \cite{Conneau}. On the other hand, there has been limited uses of transfer learning in health informatics. For example, Alaa \textit{et al.} \cite{alaa2018personalized} have used the parameters of a Gaussian expert process trained on patients with stable conditions for patients with deteriorating conditions.

In this paper, we propose a novel framework for ECG analysis that is able to represent the signal in a way that is transferable between different tasks. For this to happen, we describe a deep neural network architecture which offers a considerable capacity for learning such representations. This network has been trained on the task of arrhythmia detection for learning which it is plausible to assume that the model needs to learn most of the shape-related features of the ECG signal. Also, we have a large amount of labeled data for this task, which makes it easy to train a network with a large amount of parameters. Furthermore, we show that the signal representation learned from this task is successfully transferable to the task MI prediction using ECG signals. This method allows us to use these deep representations to share knowledge between ECG recognition tasks for which enough information may not be available for training a deep architecture.

The rest of this paper is organized as follows. Section~\ref{sec:Datasets} explains the datasets used in this study. Section~\ref{sec:Methodology} presents the proposed method. Section~\ref{sec:Results} presents results of the suggested method on different task and comparison of them with other works in the literature. Finally, Section~\ref{sec:Conclusion} concludes the paper.

\section{Datasets}
\label{sec:Datasets}
In this paper, we use PhysioNet MIT-BIH Arrhythmia and PTB Diagnostic ECG Databases as data source for labeled ECG records \cite{goldberger2000physiobank,moody2001impact,bousseljot1995nutzung}. Furthermore, we demonstrate that the knowledge learned from the former database can be successfully transferred for training inference models for the latter. In all of our experiments, we have used ECG lead II re-sampled to the sampling frequency of $125 Hz$ as the input.

The MIT-BIH dataset consists of ECG recordings from $47$ different subjects recorded at the sampling rate of $360 Hz$. Each beat is annotated by at least two cardiologists. We use annotations in this dataset to create five different beat categories in accordance with Association for the Advancement of Medical Instrumentation (AAMI) EC57 standard \cite{association1998testing}. See Table~\ref{tab:aami} for a summary of mappings between beat annotations in each category.

\begin{table}[!t]
\renewcommand{\arraystretch}{1.3}
\caption{Summary of mappings between beat annotations and AAMI EC57 \cite{association1998testing} categories.}
\label{tab:aami}
\centering
\begin{tabular}{lp{2.0in}}
\hline
\textbf{Category} & \textbf{Annotations} \\
\hline
\hline
\textbf{N} & \begin{itemize} \item Normal \item Left/Right bundle branch block \item Atrial escape \item Nodal escape \end{itemize}\\
\hline
\textbf{S} & \begin{itemize} \item Atrial premature  \item Aberrant atrial premature \item Nodal premature \item Supra-ventricular premature \end{itemize}\\
\hline
\textbf{V} & \begin{itemize} \item Premature ventricular contraction \item Ventricular escape \end{itemize}\\
\hline
\textbf{F} & \begin{itemize} \item Fusion of ventricular and normal \end{itemize}\\
\hline
\textbf{Q} & \begin{itemize}
    \item Paced \item Fusion of paced and normal \item Unclassifiable \end{itemize}\\
\hline
\end{tabular}
\end{table}

The PTB Diagnostics dataset consists of ECG records from $290$ subjects: $148$ diagnosed as MI , $52$ healthy control, and the rest are diagnosed with $7$ different disease. Each record contains ECG signals from $12$ leads sampled at the frequency of $1000 \text{Hz}$. In this study we have only used ECG lead II, and worked with MI and healthy control categories in our analyses.

\section{Methodology}
\label{sec:Methodology}

\subsection{Preprocessing}
\label{sec:Preprocessing}
\begin{figure}[!t]
\centering
\includegraphics[width=\columnwidth]{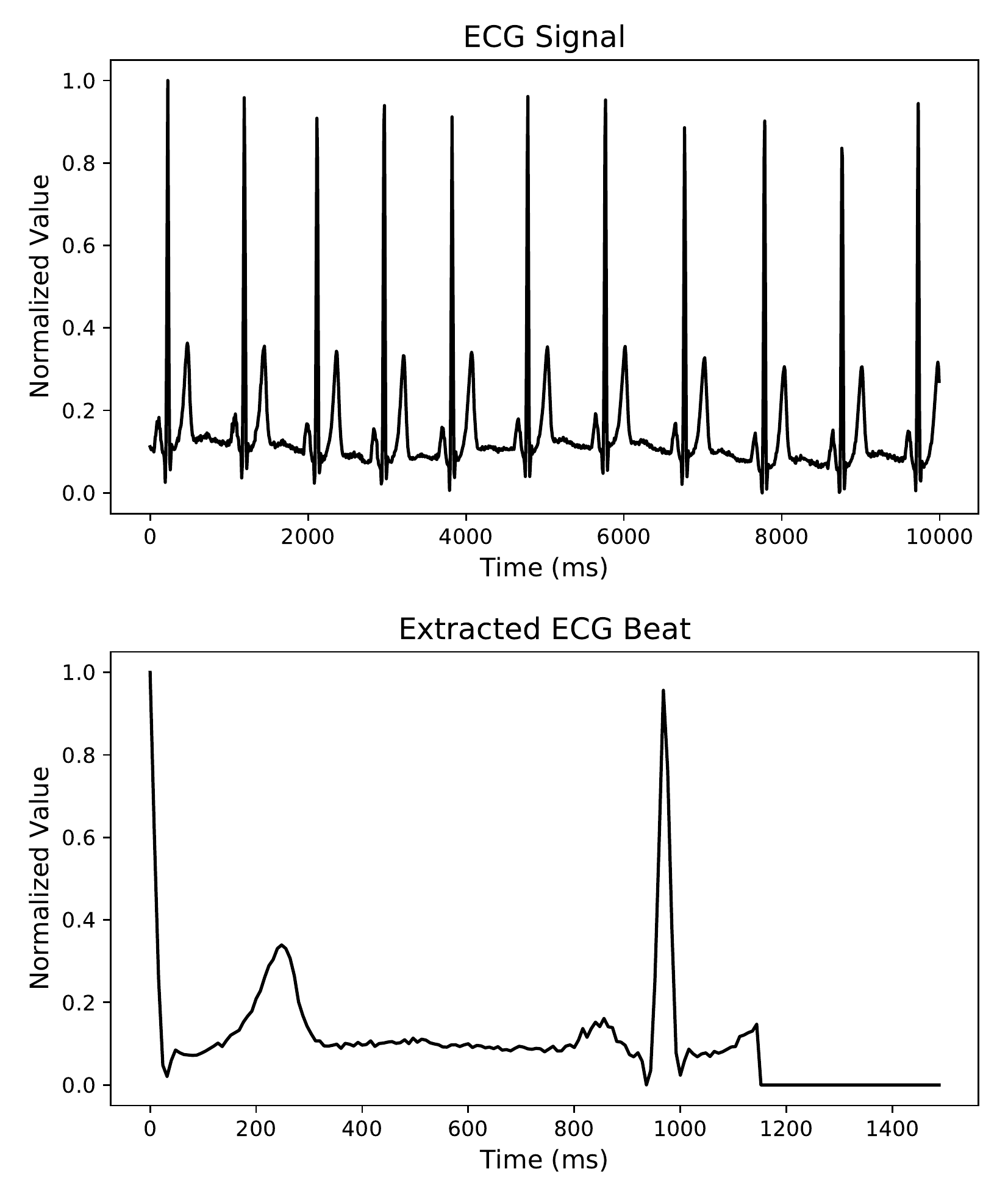}%
\caption{An example of a 10s ECG window and an extracted beat from it.}
\label{fig:beat_extraction}
\end{figure}

As ECG beats are inputs of the proposed method we suggest a simple and yet effective method for preprocessing ECG signals and extracting beats. The steps used for extracting beats from an ECG signal are as follows (see Fig.~\ref{fig:beat_extraction}):
\begin{enumerate}
    \item Splitting the continuous ECG signal to $10s$ windows and select a $10s$ window from an ECG signal.
    \item Normalizing the amplitude values to the range of between zero and one.
    \item Finding the set of all local maximums based on zero-crossings of the first derivative.
    \item Finding the set of ECG R-peak candidates by applying a threshold of $0.9$ on the normalized value of the local maximums.
    \item Finding the median of R-R time intervals as the nominal heartbeat period of that window ($T$).
    \item For each R-peak, selecting a signal part with the length equal to $1.2 T$.
    \item Padding each selected part with zeros to make its length equal to a predefined fixed length.
\end{enumerate}

It is worth mentioning that the suggested beat extraction method is simple and effective in extracting R-R intervals from signals with different morphologies. For example, we have not used any form of filtering or any processing that makes any assumption about the signal morphology or spectrum. Additionally, all the extracted beats have identical lengths which is essential for being used as inputs to the subsequent processing parts.

\subsection{Training the Arrhythmia Classifier}
\label{sec:Training the Arrhythmia Classifier}
In this paper we suggest training a convolutional neural network for classification of ECG beat types on the MIT-BIH dataset. The trained network not only can be used for the purpose of beat classification, but also in the next section we show that it can be used as an informative representation of heartbeats.

\begin{figure}[!t]
\centering
\includegraphics[width=0.8\columnwidth]{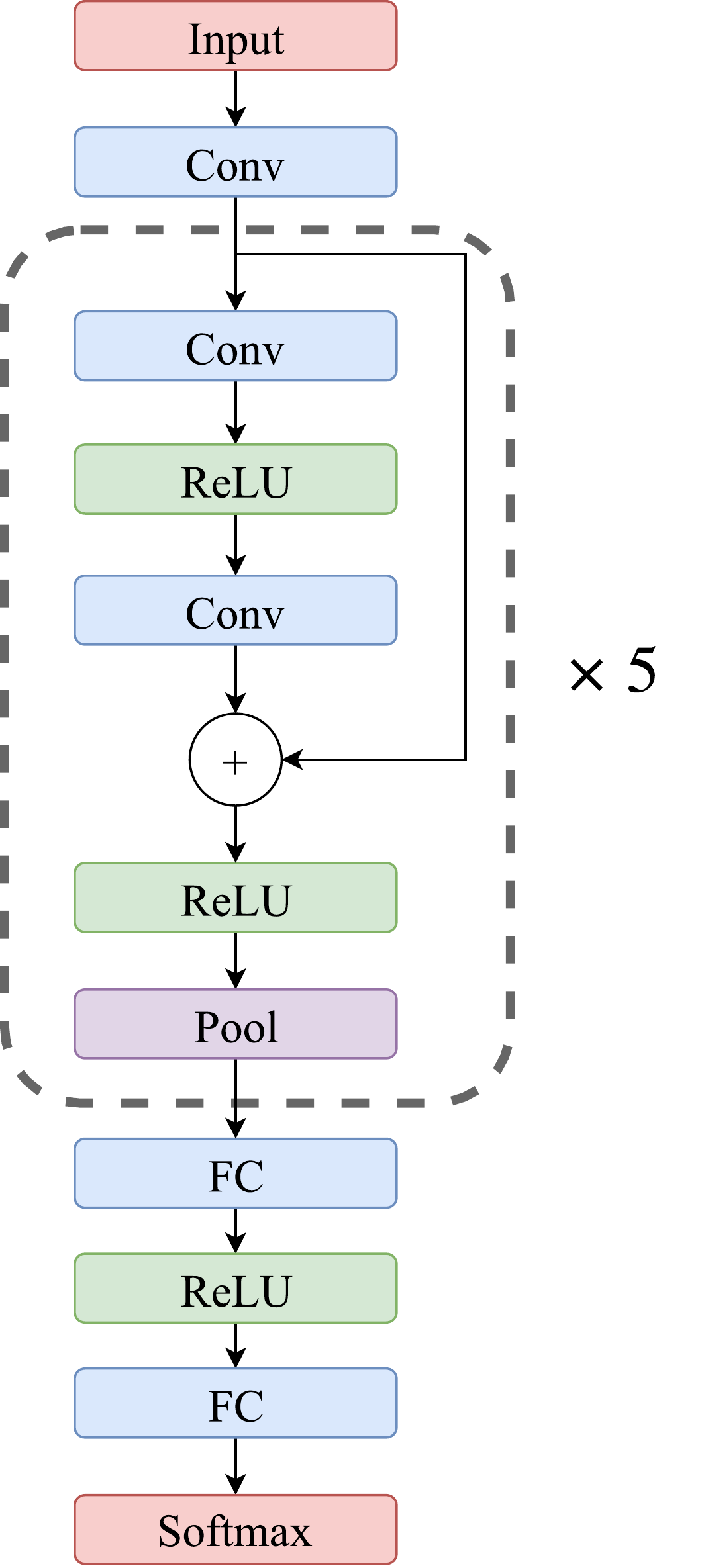}%
\caption{Architecture of the proposed network.}
\label{fig:net_mitbih}
\end{figure}

Fig.~\ref{fig:net_mitbih} illustrates the network architecture proposed for the beat classification task. Extracted beats, as explained in Section~\ref{sec:Preprocessing}, are used as inputs. Here, all convolution layers are applying 1-D convolution through time and each have $32$ kernels of size $5$. We also use max pooling of size $5$ and stride $2$ in all pooling layers. The predictor network consists of five residual blocks followed by two fully-connected layers with $32$ neurons each and a softmax layer to predict output class probabilities. Each residual block contains two convolutional layers, two ReLU nonlinearities \cite{nair2010rectified}, a residual skip connection \cite{he2016deep}, and a pooling layer. In total, the resulting network is a deep network consisting of $13$ weight layers.

\subsection{Training the MI Predictor}
After training the network suggested in Section~\ref{sec:Training the Arrhythmia Classifier}, we use the output activations of the very last convolution layer as a representation of input beats. Here, we use this representation as input to a two layer fully-connected network with $32$ neurons at each layer to predict MI. It is noteworthy to mention that during the training for the MI prediction task, we freeze the weights for all other layers aside from the last two. In other words, we only train the last two network layers and use the learned representation of Section~\ref{sec:Training the Arrhythmia Classifier}. 

\subsection{Implementation Details}
In all experiments, TensorFlow computational library \cite{abadi2016tensorflow} is used for model training and evaluation. Cross entropy loss on the softmax outputs is used as the loss function. For training the networks, we used Adam optimization method \cite{kingma2014adam} with the learning rate, beta-1, and beta-2 of $0.001$, $0.9$, and $0.999$, respectively. Learning rate is decayed exponentially with the decay factor of $0.75$ every $10000$ iterations. Training all the networks took less than two hours on a GeForce GTX 1080Ti processor.


\section{Results}
\label{sec:Results}

\begin{figure}[!t]
\centering
\includegraphics[width=\columnwidth]{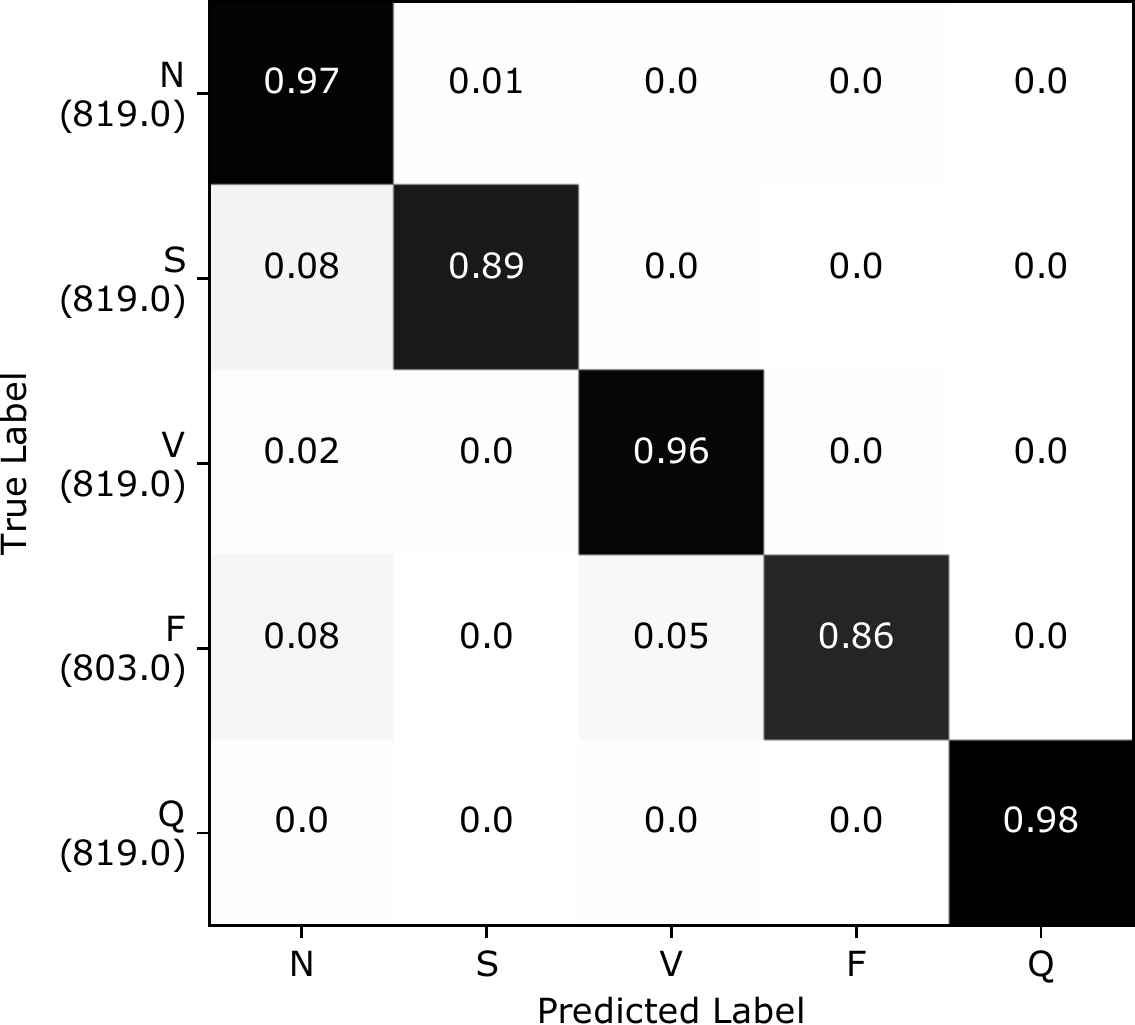}%
\caption{Confusion matrix for heartbeat classification on the test set. Total number of samples in each class is indicated inside parenthesis. Numbers inside blocks are number of samples classified in each category normalized by the total number of samples and rounded to two digits.}
\label{fig:confmat_mitbih}
\end{figure}

\subsection{Arrhythmia Classification and learning the representation}
We evaluated the arrhythmia classifier of Section~\ref{sec:Training the Arrhythmia Classifier} on $4079$ heartbeats (about $819$ from each class) that are not used in the network training phase. Note that the dataset is being augmented to reach a balance in the number of beats in each category. Fig.~\ref{fig:confmat_mitbih} presents the confusion matrix of applying the classifier on the test set. As it can be seen from this figure, the model is able to make accurate predictions and distinguish different classes.

\begin{figure*}[!t]
\centering
\subfloat[]{\includegraphics[width=\columnwidth]{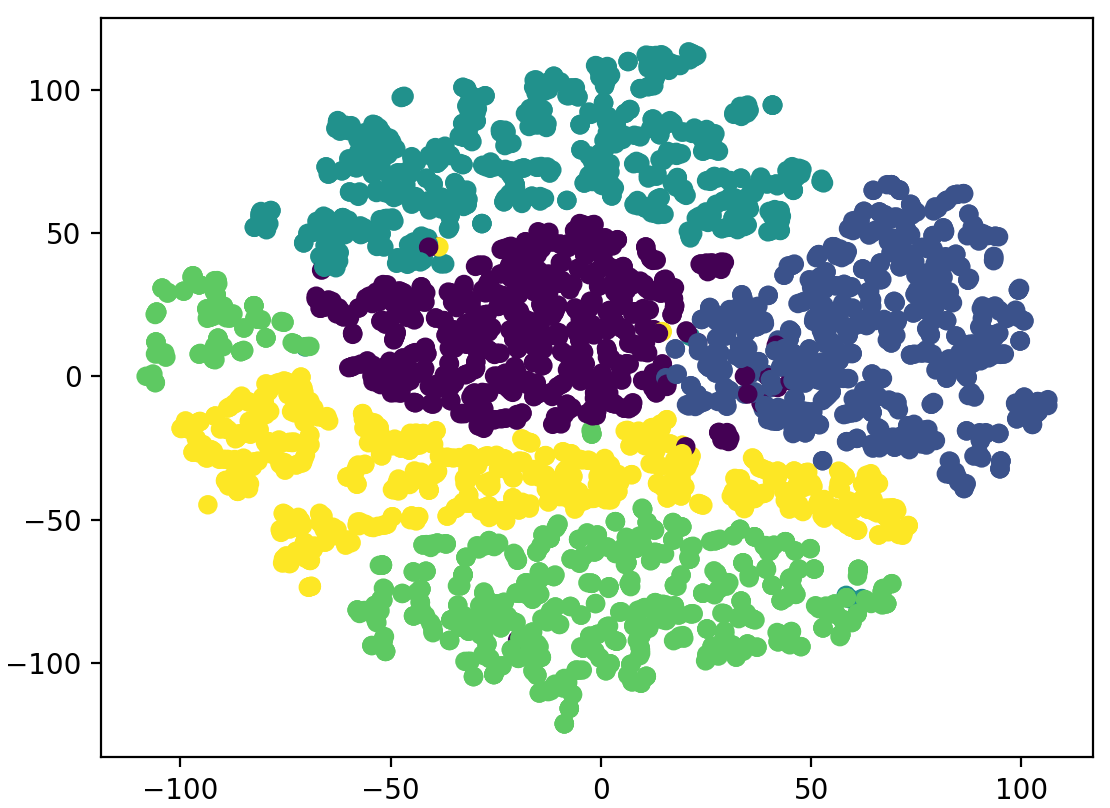}%
\label{fig:tsne_mitbih}}
\subfloat[]{\includegraphics[width=\columnwidth]{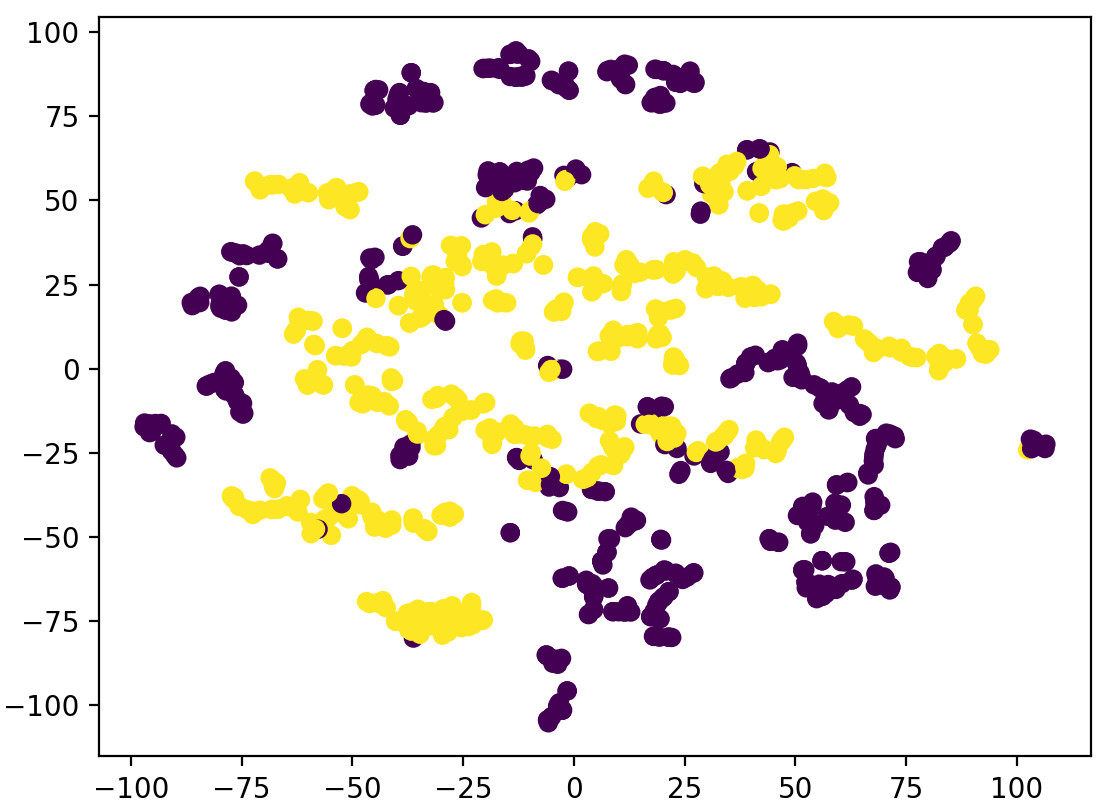}%
\label{fig:tsne_mi}}
\caption{t-SNE visualization of the learned representation: (a) samples from MIT-BIH for ECG beat classification (b) samples from PTB dataset for MI classification. Labels for each task are indicated with colors (best viewed in color).}
\label{fig:tsne}
\end{figure*}

Table~\ref{tab:results_mitbih} presents the average accuracy of the proposed method and compares it with other relevant methods in the literature. While suggesting a predictor for MIT-BIH is not the sole purpose of this study, according to the results, the accuracies achieved in this paper are competitive to the state of the art methods. The main reason behind this might be the fact that we have used residual connections in our network architecture which allows us to train deeper networks compared to using traditional convolutional architectures.

\begin{table}[!t]
\renewcommand{\arraystretch}{1.3}
\caption{Comparison of heartbeat classification results.}
\label{tab:results_mitbih}
\centering
\begin{tabular}{lcc}
\hline
\textbf{Work}  & \textbf{Approach} & \textbf{Average Accuracy (\%)} \\
\hline
\hline
\textbf{This Paper} & \textbf{Deep residual CNN} & \bm{$93.4$} \\

Acharya \textit{et al.} \cite{acharya2017deep} & Augmentation + CNN & $93.5$  \\

Martis \textit{et al.} \cite{martis2013application} & DWT + SVM & $93.8$  \\

Li \textit{et al.} \cite{li2016ecg} & DWT + random forest & $94.6$  \\

\hline
\end{tabular}
\end{table}

\subsection{MI Classification using the learned representation}
We have trained our MI predictor using the learned representations, and took $80\%$ of the PTB dataset as our training set. We have used the remaining $20\%$ to test our model. Table~\ref{tab:results_mi} presents a comparison between the average accuracy, precision, and recall of the proposed method for MI classification and other work in the literature. The performance of the proposed method is better than all other works except the method suggested by Sharma \textit{et al.} \cite{sharma2015multiscale} that reports higher accuracy and precision values. However, it noteworthy to mention that Sharma \textit{et al.} use 12-lead ECG as opposed to us using only the lead II.

\begin{table}[!t]
\renewcommand{\arraystretch}{1.3}
\caption{Comparison of MI classification results.}
\label{tab:results_mi}
\centering
\begin{tabular}{lccc}
\hline
\textbf{Work}  & \textbf{Accuracy (\%)} & \textbf{Precision (\%)} & \textbf{Recall (\%)}\\
\hline
\hline
\textbf{This Paper\footnotemark[1]} & \bm{$95.9$} & \bm{$95.2$} & \bm{$95.1$} \\

Acharya \textit{et al.} \cite{acharya2017application}\footnotemark[1] & $93.5$ & $92.8$ & $93.7$ \\ 

Safdarian \textit{et al.} \cite{safdarian2014new}\footnotemark[1] & $94.7$ & $-$ & $-$ \\ 

Kojuri \textit{et al.} \cite{kojuri2015prediction}\footnotemark[2] & $95.6$ & $97.9$ & $93.3$ \\ 

Sun \textit{et al.} \cite{sun2012ecg}\footnotemark[3] & $-$ & $82.4$ & $92.6$ \\ 

Liu \textit{et al.} \cite{liu2015novel}\footnotemark[3] & $94.4$ & $-$ & $-$ \\ 

Sharma \textit{et al.} \cite{sharma2015multiscale}\footnotemark[3] & $96$ & $99$ & $93$ \\ 

\hline
\end{tabular}
\begin{flushleft}
\footnotemark[1]: PTB dataset, ECG lead II \\
\footnotemark[2]: dataset collected by authors, 12-lead ECG \\
\footnotemark[3]: PTB dataset, 12-lead ECG \\

\end{flushleft}

\end{table}

\subsection{Visualization of the learned representation}

In order to visualize the learned representation, we have used t-SNE visualization method \cite{maaten2008visualizing} to map high-dimensional vector created by the last convolutional layer to the 2D space. In a nutshell, t-SNE creates a mapping such that the joint probability of data-points appearing close to each other in the high-dimensional space is similar to the same probability distribution in the low-dimensional mapped points. 

Fig.~\ref{fig:tsne_mitbih} illustrates the visualization of the learned representation on the MIT-BIH dataset samples. As it can be seen from this figure, data-points from different classes are easily separable using the learned representation. Fig.~\ref{fig:tsne_mi} shows the visualization of the MI classification task on the PTB samples using the representation trained on MIT-BIH. It can be inferred from this figure that the transferred representation for the beat classification task is able to provide a reasonable separation for the MI classification task. It should be noted that here we only use class labels to colorize the plots and other than this we do not use sample labels in the visualizations.


\section{Conclusion}
\label{sec:Conclusion}
In this study we have presented a method for ECG heartbeat classification based on a transferable representation. Specifically, we have trained a deep convolutional neural network with residual connections for the arrhythmia classification task and shown that the representation learned for this task can be used as a base to train accurate classifiers for the classification of MI. According to the results, the suggested method is able to make predictions on both tasks with accuracies comparable to the state of the art methods in the literature. Furthermore, we visualized the learned representation using t-SNE method and illustrated the effectiveness of the proposed approach. 




\end{document}